\documentstyle[12pt]{article}
\begin{document}

\begin{flushright}{NUP-A-2004-4}
\end{flushright}
\vskip 0.5 truecm

\begin{center}
{\Large{\bf Remark on Natural Models of Neutrinos}}
\end{center}
\vskip .5 truecm
\centerline{\bf Kazuo Fujikawa }
\vskip .4 truecm
\centerline {\it Institute of Quantum Science, College of 
Science and Technology}
\centerline {\it Nihon University, Chiyoda-ku, Tokyo 101-8308, 
Japan}
\vskip 0.5 truecm

\makeatletter
\@addtoreset{equation}{section}
\def\theequation{\thesection.\arabic{equation}}
\makeatother

\begin{abstract}
We comment on what the naturalness argument of 't Hooft implies 
for a minimal extension of the standard model which 
incorporates right-handed neutrinos with generic mass
terms. If this Lagrangian is taken as a low energy effective 
theory, the idea of  pseudo-Dirac neutrinos with very small 
masses is consistent with the naturalness argument of 't Hooft. 
This argument is based on an observation that the 
right-handed components of neutrinos in the massless limit
exhibit an extra enhanced symmetry which is absent in other 
charged fermions. This enhanced symmetry is reminiscent of  the 
Nambu-Goldstone fermions associated with spontaneously broken
supersymmetry. The conventional seesaw scenario gives  another 
natural solution if the ultra-heavy right-handed neutrinos 
are integrated out in defining a low energy effective theory.
\end{abstract}

\section{Introduction}
It appears that no reliable theory of lepton and quark masses is 
known in the standard model~\cite{weinberg}, and thus the 
fermion masses and mixing angles are purely phenomenological 
parameters at this moment.  
The small neutrino masses indicated by the oscillation 
experiments~\cite{pakvasa, smirnov} could be even more 
deceptive than 
ordinary lepton and quark masses, and it may be worth examining 
the small neutrino masses from a general perspective 
independently of explicit detailed models.

 In this note we comment on the general aspects of 
neutrino masses in the standard model with generic neutrino
mass terms added by taking the naturalness argument of 
't Hooft~\cite{'t hooft} as a guiding principle.
At this moment,  the popular picture of neutrinos with small 
observed masses appears to be the seesaw scenario~\cite{seesaw} 
where the right-handed neutrinos with
 huge masses, when integrated out, induce small masses for the 
left-handed neutrinos in the low energy effective theory. The 
small lepton number violating terms in the low energy effective 
theory are thus natural  in the sense 
of 't Hooft, since one recovers the lepton 
number conservation if one sets the small masses for the 
left-handed neutrinos to be zero.
On the other hand, the Dirac neutrinos with small masses are 
consistent with oscillation experiments\cite{pakvasa}, but such 
neutrino mass terms 
appear to be neither generic nor natural~\footnote{From the view point of chiral symmetry, fermion masses are renormalized multiplicatively and thus any values of fermion masses may be said to satisfy the naturalness condition of 't Hooft. We search
for other symmetries.}
. In this note we point 
out that almost Dirac-type neutrinos with tiny 
masses~\footnote{These neutrinos are known generically as 
pseudo-Dirac neutrinos~\cite{wolfenstein} - \cite{kobayashi}, 
and the possible 
dynamical schemes for pseudo-Dirac neutrinos have been analyzed 
from different view points in the past. See, for example, 
\cite{langacker, cleaver, chang, extradim} and references 
therein.} are consistent with the naturalness argument of 
't Hooft in the framework of a minimal extension of the 
standard model which contains right-handed neutrinos with 
generic mass terms. Our argument is based on an observation 
that the right-handed components of neutrinos exhibit an 
extra enhanced symmetry in the 
massless limit, which is absent in other charged fermions.
This enhanced symmetry turns out to be reminiscent of the 
Nambu-Goldstone fermions~\cite{volkov, salam} associated with 
spontaneously broken supersymmetry.

\section{One-generation model}

We first study  one-generation of leptons to explain the 
essence of the argument.
We consider a minimal extension of the standard 
model~\cite{weinberg} by 
incorporating the right-handed neutrino
\begin{equation}
 \psi_{L}=\left(
 \begin{array}{c}
  \nu_{L}\\ e_{L}
 \end{array}
 \right), \ \
 \psi_{R}=\left(
 \begin{array}{c}
  \nu_{R}\\ e_{R}
 \end{array}
 \right)
\end{equation}
and the part of the Lagrangian relevant to our discussion 
is given by 
\begin{eqnarray}
{\cal L}&=&\overline{\psi}_{L}i\gamma^{\mu}
(\partial_{\mu} - igT^{a}W_{\mu}^{a}
            - i\frac{1}{2}g^{\prime}Y_{L}B_{\mu})\psi_{L}
\nonumber\\
        && +\overline{\psi}_{R}i\gamma^{\mu}(\partial_{\mu}
            - i\frac{1}{2}g^{\prime}Y_{R}B_{\mu})\psi_{R}
\nonumber\\
           &&+[ - \overline{\psi}_{R}M\psi_{L}
            -\frac{1}{2}\nu_{R}^{T}C\mu\nu_{R}]+ h.c.
\end{eqnarray}
with $Y_{L}=-1$ and 
\begin{equation}
Y_{R}=\left(\begin{array}{cc}
            0&0\\
            0&-2
            \end{array}\right).
\end{equation}

The Dirac mass term in the unitary gauge is given by
\begin{equation}
M=\left(\begin{array}{cc}
            m_{D}+(m_{D}/v)\varphi&0\\
            0&m_{e}+(m_{e}/v)\varphi
            \end{array}\right)
\end{equation} 
where $v$ stands for the vacuum value of the Higgs field, and 
the variable $\varphi$ above
stands for the Higgs field after subtracting the vacuum value.
The operator $C$ stands for the charge conjugation matrix for
spinors.\footnote{We adopt the charge conjugation matrix
convention
\begin{eqnarray}
C\gamma^{\mu}C^{-1}=-(\gamma^{\mu})^{T}, \ \ C\gamma_{5}C^{-1}
=\gamma^{T}_{5},\ \ C^{\dagger}C=1,\ \ C^{T}=-C.
\nonumber
\end{eqnarray}
} The term with $\mu$ in the above Lagrangian is the 
Majorana mass term for the right-handed neutrino.

The above Lagrangian is formally the same as that
 for the conventional seesaw scenario if one chooses
$\mu^{2}\gg m^{2}_{D}$. The seesaw picture is mainly motivated 
by a grand unification idea such as $SO(10)$~\cite{fukugita}.
In the following, we take the above Lagrangian as a 
{\em low energy effective theory} and apply the naturalness 
argument of 't Hooft to it. It is then argued that the 
choice $\mu^{2}\ll m^{2}_{D}$ with $m_{D}$ much smaller than the 
charged lepton mass is also natural in the low energy
effective theory.\footnote{The right-handed neutrino $\nu_{R}$ 
with a huge mass is  not allowed to appear in low energy 
effective theory and thus it is  integrated out in the seesaw 
scenario, while $\nu_{R}$ appears in the low energy effective 
theory itself in the present scheme.}

The free part of the neutrino of this model is given by 
\begin{equation}
{\cal L}_{\nu}=\bar{\nu}i\gamma^{\mu}\partial_{\mu}\nu
-\bar{\nu}m_{D}\nu-(\frac{1}{2}\nu_{R}^{T}C\mu\nu_{R}+ h.c.).
\end{equation}
We here present an analysis of this simple model in some detail
to set a notation for the realistic three-generation model in
the next section.
 The mass terms are generic, and $m_{D}$ and $\mu$ are chosen to 
be real by a suitable choice of the phase convention of field 
variables in the case of a single flavor. By defining 
$N_{L}^{T}C=\bar{\nu}_{R}$ (and consequently 
$\nu_{R}^{T}C=\bar{N}_{L}$), 
the above Lagrangian is written as  
\begin{eqnarray}
{\cal L}_{\nu}&=&\bar{\Psi}_{L}i\gamma^{\mu}\partial_{\mu}
\Psi_{L} - \frac{1}{2}\Psi^{T}_{L}C{\cal M}\Psi_{L} + h.c. 
\end{eqnarray}
for the neutrino field defined by 
\begin{eqnarray}
\Psi_{L}&=&\left(\begin{array}{c}
          \nu_{L}\\
          N_{L}\\
          \end{array}\right)
\end{eqnarray}
with the mass matrix 
\begin{eqnarray}
{\cal M}&=&\left(\begin{array}{cc}
          0&m^{T}_{D}\\
          m_{D}&\mu\\
          \end{array}\right),
\end{eqnarray}
though $m^{T}_{D}=m_{D}$ in the present single flavor case.
We thus have the Majorana mass eigenvalues
\begin{equation}
m_{\pm}=\frac{1}{2}\mu \pm 
\sqrt{m_{D}^{2}+(\frac{1}{2}\mu)^{2}}
\end{equation}
and the mixing angle between $\nu_{L}$ and $N_{L}$, which 
defines an orthogonal transformation,  is given by 
\begin{equation}
\tan \theta =\frac{m_{D}}{\mu+\sqrt{m_{D}^{2}+\mu^{2}}}.
\end{equation}
The corresponding neutrino mass eigenstates are given by
\begin{eqnarray}
\Psi_{L}\rightarrow \left(\begin{array}{c}
          \nu_{(1)L}\\
          \nu_{(2)L}\\
          \end{array}\right)
          = \left(\begin{array}{c}
          \cos\theta\nu_{L}+\sin\theta N_{L}\\
          -\sin\theta\nu_{L}+\cos\theta N_{L}\\
          \end{array}\right).
\end{eqnarray}
The mixing angle $\theta$ is close to $\pi/4$
for the case of $m_{D}\gg \mu$.
The Majorana mass eigenstates are then defined by 
\begin{eqnarray}
\Psi_{M}  = \left(\begin{array}{c}
          \nu_{(1)L}+\nu_{(1)R}\\
          \nu_{(2)L}+\nu_{(2)R}\\
          \end{array}\right)
          \equiv \left(\begin{array}{c}
          \nu_{(1)M}\\
          \nu_{(2)M}\\
          \end{array}\right)
\end{eqnarray}
with $\nu_{(1)R}=[\bar{\nu}_{(1)L}C^{-1}]^{T}$ and 
$\nu_{(2)R}=[\bar{\nu}_{(2)L}C^{-1}]^{T}$,
and the Lagrangian (2.5) is written as 
\begin{eqnarray}
{\cal L}_{\nu}=\frac{1}{2}\Psi^{T}_{M}Ci\gamma^{\mu}
\partial_{\mu}\Psi_{M} 
-\frac{1}{2}\Psi^{T}_{M}C\left(\begin{array}{cc}
          m_{+}&0\\
          0&m_{-}\\
          \end{array}\right)\Psi_{M}.
\end{eqnarray}
The minus sign in the neutrino mass is taken care of by a 
suitable chiral transformation. Effectively, the above may 
be regarded as a decomposition of a single Dirac neutrino,
which is specified by $\nu_{L}$ and $N_{L}$, to
a linear combination of two massive Majorana neutrinos.

The weak  interaction is described by 
\begin{eqnarray}
\nu_{L}=\cos\theta\nu_{(1)L}-\sin\theta\nu_{(2)L}
=(\frac{1-\gamma_{5}}{2})(\cos\theta\nu_{(1)M}
-\sin\theta\nu_{(2)M})
\end{eqnarray}
and the weak singlet state is given by 
\begin{eqnarray}
N_{L}=\sin\theta\nu_{(1)L}+\cos\theta\nu_{(2)L}
=(\frac{1-\gamma_{5}}{2})(\sin\theta\nu_{(1)M}
+\cos\theta\nu_{(2)M}).
\end{eqnarray}
The neutrino oscillation~\cite{MNS, gribov, bilenky2} in this 
case is 
similar to the (Dirac) neutrino rotation in a strong magnetic 
field\cite{FS}\cite{wolfenstein}; the weak active
left-handed state $\nu_{L}$ is rotated to a weak in-active 
right-handed state $N_{L}=[\bar{\nu}_{R}C^{-1}]^{T}$.

We now argue that  
\begin{equation}
\mu^{2}\ll m^{2}_{D}
\end{equation}
is consistent with the naturalness of 't Hooft. The basic 
postulate of the naturalness of 't Hooft is that a small 
parameter in low energy effective theory is natural only when 
one obtains 
an enhanced symmetry by setting such a small parameter 
to be 0~\cite{'t hooft}. In the present case, if one sets
$\mu=0$ one recovers an extra enhanced symmetry, namely, the 
fermion number symmetry; this argument in the context of  
low-energy effective theory is a standard one in any analysis 
of pseudo-Dirac neutrinos~\cite{wolfenstein}-\cite{extradim}. 
We thus have the natural Majorana masses of the neutrinos as
\begin{equation}
m_{\pm}\simeq\frac{1}{2}\mu\pm m_{D}
\end{equation}
where the negative mass is made positive by a suitable chiral
transformation. 

We next argue that the value of $m_{D}$ which is much smaller 
than other charged lepton and quark masses is consistent with 
the naturalness of 't Hooft, since if one sets $m_{D}=0$ in 
(2.4) (with $\mu=0$) one finds an extra enhanced symmetry
\begin{equation}
\nu_{R}(x)\rightarrow \nu_{R}(x)+\eta_{R}
\end{equation}
where $\eta_{R}$ is a constant spinor, or in the Majorana
notation
\begin{equation}
\psi_{M}(x)\rightarrow \psi_{M}(x)+\eta_{M}
\end{equation}
where
\begin{equation}
\psi_{M}(x)=\nu_{R}(x)+\nu_{L}(x)
\end{equation}
with $\nu_{L}=[\bar{\nu}_{R}C^{-1}]^{T}$.
The existence of this special symmetry is a result of the fact 
that only $\nu_{R}$ is gauge singlet in the standard model.
To adopt this simple symmetry as a basic symmetry
of the effective theory, we need to assume that the enhanced 
symmetry (2.18) is a basic symmetry of the  theory underlying 
the standard model. In more technical terms, we need to 
assume that the right-handed component of the neutrino couples 
to other heavier degrees of freedom in the effective Lagrangian,
 which do not explicitly appear in the standard model,  either 
through the coupling which is proportional to the neutrino mass 
or through the derivative coupling which is suppressed in the 
low energy effective theory. In this note we assume that this 
is the case, though we have no convincing reason to assert why 
it should be so.

In the limit where all the Dirac-type masses vanish with the 
vacuum value $v$ kept fixed (and with $\mu=0$), all the 
fermions in the standard model become chirally symmetric. But 
only the right-handed component of the neutrino 
has the above extra stronger symmetry. Our assumption in
this note is that this enhanced symmetry (2.18)
is more effective for the tiny neutrino mass than the chiral 
symmetry which affects all the fermion masses universally.

Our analysis so far has no connection with supersymmetry.
However, the symmetry  (2.18) where the fermion field 
transforms inhomogeneously with a constant component is 
reminiscent of a Nambu-Goldstone fermion in spontaneously broken
 supersymmetric theory~\cite{wess}. Since the Nambu-Goldstone 
fermion implies a tiny mass, it may be tempting to entertain 
the idea of possible connection of the enhanced symmetry (2.18) 
with supersymmetry. We however note that the Nambu-Goldstone 
fermions satisfy the enhanced symmetry (2.18) in low energy 
scattering amplitudes~\cite{wess} 
(and thus in the low energy efffective action), but the 
enhanced symmetry (2.18) by itself does not necessarily imply 
the existence of spontaneously broken supersymmetry in the deep 
level. \footnote{If our speculation on 
the possible connection with supersymmetry should be  valid, the 
transformation law (2.18) would be  replaced by 
\begin{eqnarray}
\nu_{R}(x)\rightarrow \nu_{R}(x)+\eta_{R}+O_{R}(x)
\nonumber
\end{eqnarray}
in a full renormalizable supersymmetric theory, where $O_{R}(x)$ 
stands for the fields or composite operators representing the 
heavy degrees of freedom which consist of superpartners.  
The observed  tiny neutrino mass 
and the absence of supersymmetric particles in the standard 
model suggest that (presumed) supersymmetry is explicitly 
broken in a very specific way such that the notion of 
Nambu-Goldstone fermions is not completely spoiled.}

It may be interesting to recall that the idea of the neutrino as 
a Nambu-Goldstone fermion was suggested  immediately after the 
discovery of supersymmetry~\cite{volkov, salam}, but the 
idea has 
been later abandoned since the (left-handed) neutrino in the 
Fermi interaction does not decouple in the low-energy limit
contrary to the basic property expected for a Nambu-Goldstone 
particle~\cite{dewit}.
In contrast, the right-handed component of the neutrino in fact
decouples from the Fermi interaction in the low energy 
limit.\footnote{The right-handed component of the neutrino in 
the massless limit simply decouples from all the interactions 
in the standard model. The decoupling of the right-handed 
neutrino from possible heavy superpartners in  low energy 
effective theory  is at least consistent with the suggested 
Nambu-Goldstone nature of the right-handed neutrino.}

In any case, the present naturalness argument which is based on 
the enhanced special symmetry (2.18) for the right-handed 
neutrino does not contradict the observed fact that the neutrino
 masses are very small compared to other charged fermion masses,
regardless of whether the above enhanced symmetry 
is possibly associated with supersymmetry or not. 

\section{Three-generation model}

We now discuss a realistic three-generation model where 
three right-handed neutrinos with generic mass terms are added 
to the standard model.
Our formulas (2.1) and (2.2) are 
valid for the case of three generations of fermions if one 
understands that $e_{L,R}(x)$ there contains 3 components as 
\begin{eqnarray}
e_{L,R}(x)&\rightarrow&\left(\begin{array}{c}
          e_{L,R}(x)\\
          \mu_{L,R}(x)\\
          \tau_{L,R}(x)\\
          \end{array}\right)
\end{eqnarray} 
and, correspondingly, the neutrino fields $\nu_{L}$ and 
$\nu_{R}$ respectively contain 3 fields.
The neutrino field in (2.7) is then replaced by
\begin{eqnarray}
\Psi_{L}&=&\left(\begin{array}{c}
          \nu_{L}\\
          N_{L}\\
          \end{array}\right)
\end{eqnarray}
where
\begin{eqnarray}
\nu_{L}&=&\left(\begin{array}{c}
          \nu^{(1)}_{L}\\
          \nu^{(2)}_{L}\\
          \nu^{(3)}_{L}\\
          \end{array}\right), \ \  
N_{L}=\left(\begin{array}{c}
          N^{(1)}_{L}\\
          N^{(2)}_{L}\\
          N^{(3)}_{L}\\
          \end{array}\right)
     =\left(\begin{array}{c}
          [\bar{\nu}^{(1)}_{R}C^{-1}]^{T}\\
          {[\bar{\nu}^{(2)}_{R}C^{-1}]}^{T}\\
          {[\bar{\nu}^{(3)}_{R}C^{-1}]}^{T}\\
          \end{array}\right).
\end{eqnarray}
The extra fields $N^{(1)}_{L}\sim N^{(3)}_{L}$ stand for the 
gauge singlet neutrinos.
The mass matrix of the neutrinos in (2.8) is replaced by 
\begin{eqnarray}
{\cal M}&=&\left(\begin{array}{cc}
          0&m^{T}_{D}\\
          m_{D}&\mu\\
          \end{array}\right)
\end{eqnarray}
where $m_{D}$ and $\mu$ now stand for $3\times3$ matrices of 
complex numbers in general, and $\mu$ is a symmetric matrix.  
The matrix ${\cal M}$ thus contains 15 complex parameters in 
general. 

In the present three-generation case also, the naturalness 
argument of 't Hooft in low energy effective theory is 
consistent with\footnote{The notation 
$||\mu||\ll ||m_{D}||$ may be understood as meaning that
the largest eigenvalue of $\mu$ is much smaller than the 
smallest eigenvalue of $m_{D}$.} 
$||\mu||\ll ||m_{D}||$ 
because of the breaking of lepton number symmetry by the 
Majorana mass $\mu$. A further naturalness argument on the basis
of a generalization of the enhanced symmetry (2.18) for 
right-handed neutrinos, which appears if one sets $m_{D}=0$ 
(with $\mu=0$), is also consistent with the Dirac-type neutrino 
masses $m_{D}$ which are much smaller than the masses of  other 
charged leptons and quarks. We are of course assuming that 
the enhanced symmetry (2.18), due to the reasons not understood 
at this moment, is more effective for ensuring the 
tiny neutrino masses than the chiral 
symmetry which appears universally for all the fermions when 
the Dirac-type masses are set to zero.
 To be more explicit, we have a generalization of the enhanced 
symmetry (2.19) for $m_{D}=0$ (with $\mu=0$),
\begin{equation}
\psi^{(i)}_{M}(x) \rightarrow \psi^{(i)}_{M}(x) + 
\eta^{(i)}_{M},\ \ \ \ \ i=e, \mu, \tau
\end{equation}
with
$\psi^{(i)}_{M}(x)=\nu^{(i)}_{R}(x)
+[\bar{\nu}^{(i)}_{R}(x)C^{-1}]^{T}$.  
The existence of this special symmetry is a result of the fact 
that only $\nu_{R}$ is gauge singlet in the standard model.
Our basic assumption is that the above enhanced symmetry (3.5)
is a basic symmetry of the theory underlying the standard model.

Our analysis so far is independent of supersymmetry or any other 
fermionic symmetry. However, the enhanced symmetry (3.5) is 
suggestive of the Nambu-Goldstone nature of the right-handed 
neutrinos, and
it is tempting to entertain the idea that this enhanced 
symmetry is associated with supersymmetry. But we have extra 
complications in the attempt of this interpretation in the 
three-generation case. First of all, the number of 
Nambu-Goldstone fermions 
agrees with the number of generators of spontaneously broken 
supersymmetry. The three Nambu-Goldstone fermions thus suggest 
an extended $N=3$ supersymmetry with three Majorana-type 
supercharges. The possible association of three generations of 
fermions with $N=3$ generators of supersymmetry is interesting, 
but the extended $N=3$ supersymmetry introduces a plethora of 
exotic particles in the full theory~\cite{wess}. Also, the 
interplay between the explicit and spontaneous breakings of 
possible supersymmetry becomes more involved. 

However, we would like to indicate that the above possible 
association with supersymmetry does not lead to an outright 
contradiction, by recalling a model~\cite{bardeen} where all 
the leptons and quarks in the standard model are understood as 
Nambu-Goldstone fermions arising from spontaneously broken 
supersymmetry. In their scheme, the finite masses of leptons and
 quarks are mainly attributed to the explicit 
supersymmetry breaking by gauge interactions appearing in the 
standard model. In this interpretation, the neutrino masses
are expected to be special since the right-handed components 
are gauge singlet. In our naturalness argument above, the 
enhanced symmetry (3.5) also arises from the fact that the 
right-handed neutrinos are gauge singlet. 
Their model may show that the possible association of the 
enhanced symmetry (3.5) with supersymmetry does not lead to
an outright contradiction, though our argument here is not 
identical to their explicit model. It may be interesting to
investigate the model in~\cite{bardeen} further in view of 
our observation of the enhanced symmetry.
\\

We consider that our argument on the basis of the naturalness of
 't Hooft for the small Dirac-type neutrino masses $m_{D}$ is 
valid to the extent that the above enhanced symmetry (3.5) is 
taken to be a basic symmetry of the theory underlying the 
standard model, quite independently of the possible association 
of the enhanced symmetry (3.5) with supersymmetry. We briefly 
comment on the phenomenological implications of pseudo-Dirac 
neutrinos in the following with the proviso that this is the 
case.
 
It is known that there are 12 independent
{\em complex} parameters to characterize the lepton mixing
\cite{schechter} in the present case.
In terms of these 12 complex parameters together with 6 real 
Majorana-type  mass parameters (which comprise 15 complex 
parameters), we have the following main physical properties to 
be analyzed:
\\
1. Neutrino oscillation\\
2. Neutrinoless double $\beta$ decay\\
3. Magnetic moment\\
4. CP violation
\\
In the most general case, the analyses of these properties are
quite involved. If one assumes the limit $||\mu||\ll 
||m_{D}||$ as in our case, the analyses become slightly 
easier, but the inter-connection of the above properties is 
still complicated and interesting.      

There exist the detailed analyses of the experimental 
implications of pseudo-Dirac neutrinos in the literature, for 
example, in~\cite{beacom} and we here briefly comment on some 
limiting cases.
We classify the possible cases into several categories:\\
(i) The pure Dirac case $\mu=0$\\
In this case, the neutrino oscillation is just a standard 
one\cite{MNS, gribov}, 
and no neutrinoless double $\beta$ decay\cite{doi}. The 
magnetic moment is 
also the standard one\cite{FS}, and the CP violation is 
described by a copy of the KM scheme in the quark 
sector\cite{KM}.
\\
\\
(ii) Generic Dirac mass term $m_{D}$ with $||m_{D}||
\gg ||\mu||$\\
In this case, the mass splitting of neutrinos are mainly 
controlled by the mass eigenvalues of $m_{D}$.
The mass spectrum indicated by oscillation experiments is 
basically specified 
by $m_{D}$, since oscillation experiments indicate that the 
oscillation into ``sterile'' components $N_{L}$ is 
small~\cite{pakvasa, smirnov}. The major structure of the 
neutrino mass spectrum is determined by $m_{D}$, and each mass 
eigenstate of $m_{D}$ is modulated by the perturbation of $\mu$ 
as in the case of the single generation model.  
The double $\beta$ decay, which is induced by $\mu$, is thus much
suppressed compared to what one expects for purely Majorana 
fermions such as in the seesaw scenario. The CP violation is 
also 
basically controlled by the 
mass matrix $m_{D}$ and thus similar to the case of quarks.
The mass eigenstates of  all the neutrinos are Majorana-type
in a strict sense for $\mu\neq 0$, and thus the magnetic 
moments of neutrinos are 
basically transitional ones\cite{nieves}. However, if the 
magnitudes of 
the magnetic moments are small as is indicated by the 
calculation in the standard model\cite{FS}, the magnetic 
transition is 
mainly intra-flavor transitions. For example, the electron-type 
neutrino stays electron-type, though the magnetic transition 
may cause a transition from one electron-type Majorana 
neutrino to another electron-type Majorana neutrino. In this 
sense, the magnetic transition is similar to the case 
of the Dirac neutrinos (and the physical effects of the magnetic
transition may be rather 
minor compared to the oscillation). If the magnetic moment is 
large as is allowed by a phenomenological analysis of the
 presently available  experimental limit and if the 
magnetic field is strong enough (such as in the neighborhood of
 some of the neutron stars)\cite{FS}, the magnetic transition 
can cause 
the inter-flavor transitions. In such a case, the interplay of 
the magnetic transition and the oscillation can cause 
interesting observable phenomena\cite{LM}. 
\\
\\
(iii) Degenerate Dirac mass matrix $m_{D}$ with $||m_{D}||
\gg ||\mu||$\\
It is interesting to examine a specific limiting case where
the Dirac mass matrix after diagonalization has eigenvalues such
 as 
\begin{eqnarray}
m_{D}&=&\left(\begin{array}{ccc}
          m_{1}&0&0\\
          0&m_{1}&0\\
          0&0&m_{3}\\
          \end{array}\right),
\end{eqnarray}
with $m_{1}> m_{3}$. 

The CP violation in this limiting 
model does not arise from the mass matrix 
$m_{D}$ (since we are effectively dealing with a two-generation
model) and thus CP violation entirely comes from the CP 
violation in the lepton number violating Majorana mass term 
$\mu$. However, the effects of CP violation may not necessarily 
be small in the present degenerate case. Also the mass splitting
 among the heavier neutrinos measured by oscillation experiments
faithfully indicates the magnitude of the Majorana
mass term $\mu$, and thus the double $\beta$ dacay can take 
place at the rate estimated on the basis of the $\nu_{e}$ 
oscillation experiments.    

To be more specific, we treat the Majorana mass term 
proportional to $\mu$ as a small perturbation.          
Then the diagonalization of the Dirac mass by neglecting the 
Majorana mass for the moment gives rise to the mixing matrix 
of the left-handed neutrinos coupled to the charged weak current
 as 
\begin{eqnarray}
U&=&\left(\begin{array}{ccc}
          1&0&0\\
          0&c_{23}&s_{23}\\
          0&-s_{23}&c_{23}\\
          \end{array}\right),
\end{eqnarray}
where $c_{ij}=\cos\theta_{ij}$ and $s_{ij}=\sin\theta_{ij}$.
We used the degeneracy of $\nu_{1}$ and $\nu_{2}$ to define 
that only $\nu_{2}$ to be mixed with $\nu_{3}$. In this 
procedure the Majorana mass $\mu$ for the right-handed neutrinos
 is generally replaced by a symmetric $\tilde{\mu}$. By 
recalling our 
assumption $||m_{D}||\gg||\tilde{\mu}||\sim ||\mu||$, we 
consider that the isolated
$\nu_{3}$ is not much influenced by the Majorana mass term 
$\tilde{\mu}$.
We thus retain  only the first $2\times 2$ components of 
$\tilde{\mu}$ and analyze their effects on the degenerate 
$\nu_{1}$ and $\nu_{2}$; we let $\nu_{3}$ remain a Dirac 
neutrino in this procedure. The symmetric complex $2\times 2$ 
matrix $\tilde{\mu}$, which contains 6 real parameters, can be 
diagonalized by a $2\times 2$ unitary matrix as\cite{schechter} 
\begin{eqnarray}
\tilde{\mu}&=&u^{T}\left(\begin{array}{cc}
          \mu_{1}&0\\
          0&\mu_{2}\\
          \end{array}\right)u,
\end{eqnarray} 
where $u$ is a $2\times 2$ unitary matrix which contains 4
real parameters.

We thus have the (approximate) neutrino mixing matrix for the 
charged current
\begin{eqnarray}
U&=&\left(\begin{array}{ccc}
          1&0&0\\
          0&c_{23}&s_{23}\\
          0&-s_{23}&c_{23}\\
          \end{array}\right)\times
          \left(\begin{array}{cc}
          u^{\dagger}&0\\
          0&1\\
          \end{array}\right)
\end{eqnarray}
where the three neutrinos are still treated  as Dirac 
neutrinos.
One can confirm that the overall phase of $u$ is eliminated
and thus $u$ becomes $SU(2)$ matrix. 

The mass matrix of the first two generations of neutrinos,
which contain 4 Majorana neutrinos, is then approximately given 
by
\begin{eqnarray}
{\cal M}&=&\left(\begin{array}{cccc}
          0&0&m_{1}&0\\
          0&0&0&m_{1}\\
          m_{1}&0&\mu_{1}&0\\
          0&m_{1}&0&\mu_{2}\\
          \end{array}\right).
\end{eqnarray} 
This mass matrix is diagonalized by a further  orthogonal 
$O(4)$ matrix, and  the mass eigenvalues for 4 Majorana 
neutrinos are given by 
\begin{eqnarray}
\lambda&\simeq& m_{1}\pm \frac{\mu_{1}}{2}, \ \ \ \ 
m_{1}\pm \frac{\mu_{2}}{2}
\end{eqnarray}
after a suitable chiral transformation to make all the masses
positive. The orthogonal $O(4)$ matrix corresponds to the $O(2)$
mixing matrix in (2.11).

A salient feature of this limiting 
case is that both of the oscillation and double $\beta$ decay
are controlled by the Majorana mass term $\mu$. 
The experiments indicate that the transition to the 
``sterile'' components $N_{L}$ is small, and the 
oscillation among different flavors is 
dominant~\cite{pakvasa, smirnov}. In the present limiting case,
the oscillation between different flavors and the oscillation 
between the active and ``sterile'' neutrinos are expected to be
comparable as is indicated by the mass formula (3.10) and (3.11).
This suggests that the present limiting case (which is sometimes 
referred to as inverted hierarchy), though 
theoretically interesting, is not favored by experiments.

In the present scheme of three pseudo-Dirac neutrinos, the 
generic case (ii) appears to be most favored by 
experiments. See also \cite{beacom}.

\section{Discussion}

The Dirac-type neutrinos with tiny masses, as suggested by 
J. Steinberger~\cite{glashow} among others, are interesting but 
may appear to be neither generic nor natural.
In the present note, we started with the generic mass terms for 
three generations of leptons in a minimal extension of the 
standard model and we first argued that the neutrino mass 
matrix which is close to the Dirac mass, namely 
$\mu^{2}\ll m^{2}_{D}$, is consistent with the naturalness 
argument of 't Hooft for a low energy effective theory.
By applying a  further argument of the naturalness on the basis 
of the enhanced symmetry (3.5), we argued that the Dirac-type 
masses $m_{D}$ of the neutrinos which are much smaller than 
other lepton and quark masses in the standard model are natural.
 We also entertained the idea that this enhanced symmetry may 
possibly be related to supersymmetry in the deep level.   
 Starting with the Lagrangian which has apparently the same 
form  as in the seesaw model, we thus identify
two completely different natural models of neutrinos depending 
on how to understand the mass terms of right-handed 
neutrinos in low energy effective theory.  

The naturalness argument as such cannot be water-tight, but the
naturalness is useful in helping to guess the plausible dynamics
 behind the experimentally observed facts. Our naturalness 
argument, though may not yet be a convincing one, suggests 
a possible association of the tiny neutrino masses with the 
interesting idea of 
Nambu-Goldstone fermions~\cite{volkov, salam}. Quite 
apart  from the present analysis, it may be natural to 
expect that supersymmetry plays an intrinsic role in 
understanding the observed small neutrino masses,~\footnote{The 
neutrino masses in connection with the R-symmetry in 
supersymmetric theory have 
been analyzed in the past~\cite{aulakh, hall, chikashige, 
pakvasa, smirnov}. The properties of the theory with 
spontaneously broken supersymmetry are known to be very 
restrictive~\cite{wess}, and it may be interesting to 
examine if a concrete model, which incorporates the possible 
association of tiny neutrino masses with Nambu-Goldstone
fermions, is constructed. The idea such as supersymmetric
 D-branes or other dynamical schemes may play a central role. 
The 
spontaneous breaking of internal symmetry in supersymmetric 
solitons and the quasi Nambu-Goldstone 
fermions~\cite{buchmuller} appearing there have been analyzed 
in~\cite{nitta}, for example, but to our knowledge the modern  
analysis of Nambu-Goldstone fermions themselves with model 
building in mind appears to be missing except for the attempt 
in~\cite{bardeen}.} if supersymmetry should be realized in 
nature at all,
\\
  
I thank R. Shrock for a comment at the early stage of the 
present study.


\begin{thebibliography}{99}
\bibitem{weinberg}
S. Weinberg, Phys. Rev. Lett. {\bf 19}, 1264 (1967).\\
A. Salam, in {\em Elementary Particle Theory},
edited by N. Svartholm (Stockholm, 1968) p.367.\\
S.L. Glashow, Nucl. Phys. {\bf 22}, 579 (1961).
\bibitem{pakvasa}
S. Pakvasa and J.W.F. Valle, ``Neutrino properties
before and after KamLAND'', Proc. Indian  Natl. Sci. Acad.
{\bf 70A}, 189 (2004), and references therein.
\bibitem{smirnov}
A.Y. Smirnov, ``Neutrino physics: Open theoretical questions'',
Int. J. Mod. Phys. {\bf A19}, 1180 (2004), and references 
therein.
\bibitem{'t hooft}
G. 't Hooft, ``Naturalness, chiral symmetry, and spontaneous
chiral symmetry breaking'', in {\em Recent Developments in 
Gauge Theories}, Cargese 1979, eds. G. 't Hooft et al., (Plenum,
New York, 1990).
\bibitem{seesaw}
T. Yanagida, in Proceedings of Workshop on Unified Theory
and Baryon Number in the Universe, ed. by O. Sawada and 
A. Sugamoto (KEK report 79-18, 1979) p. 95.\\
M. Gell-Mann, P. Ramond and R. Slansky, in {\it Supergravity}, 
ed. by P. van Nieuwenhuizen and D.Z. Freedman (North-Holland, 
Amsterdam, 1979), p. 315.
\bibitem{wolfenstein}
L. Wolfenstein, Nucl. Phys. {\bf B186}, 147 (1981).
\bibitem{petcov}
S.T. Petcov, Phys. Lett. {\bf B110}, 245(1982).
\bibitem{bilenky}
S.M. Bilenky and B. Pontecorvo, Sov. J. Nucl. Phys. {\bf 38},
248 (1983).
\bibitem{bilenky2}
S.M. Bilenky and S.T. Petcov, Rev. Mod. Phys. {\bf 59}, 671 
(1987), and references therein.
\bibitem{giunti}
C. Giunti, C.W. Kim and U.W. Lee, Phys. Rev. {\bf D46}, 3034 
(1992).
\bibitem{bowes}
J.P. Bowes and R.R. Volkas, J. Phys. {\bf G24}, 1249 (1998).
\bibitem{balaji}
K.R. Balaji, A. Kalliomaki and J. Maalampi, Phys. Lett. 
{\bf B524}, 153 (2002).
\bibitem{kobayashi}
M. Kobayashi and C.S. Lim, Phys. Rev. {\bf D64}, 013003 (2001).
\bibitem{langacker}
P. Langacker, Phys. Rev. {\bf D58}, 093017 (1998).
\bibitem{cleaver}
G. Cleaver, M. Cvetic, J.R. Espinosa, L. Everett, and 
P. Langacker, Phys. Rev. {\bf D57}, 2701 (1998)
\bibitem{chang}
D. Chang and O.C. Kong, Phys. Lett. {\bf B477}, 416 (2000).
\bibitem{extradim}
K.R. Dienes, E. Dunes and T. Gherghetta, Nucl. Phys. {\bf B557},
25 (1999).\\ 
G.R. Devali and A.Y. Smirnov, Nucl. Phys. {\bf B563},
63 (1999).\\
N. Arkani-Hamed, S. Dimopoulos, G.R. Dvali and J. March-Russel,
Phys. Rev. {\bf D65}, 024032 (2002).  
\bibitem{volkov}
D.V. Volkov and V.P. Akulov, Phys. Lett. {\bf B46}, 109 (1973).
\bibitem{salam}
A. Salam and J. Strathdee, Phys. Lett. {\bf B49}, 465 (1974). 
\bibitem{dewit}
B. de Wit and D.Z. Freedman, Phys. Rev. Lett. {\bf 35}, 827 
(1975).
\bibitem{fukugita}
M. Fukugita and T. Yanagida, {\it Physics of Neutrinos and 
Applications to Astrophysics} (Springer-Verlag, Berlin
Heidelberg, 2003).
\bibitem{bardeen}
W.A. Bardeen and V. Visnjic, Nucl. Phys. {\bf B194}, 422 (1982). 
\bibitem{beacom}
J.F. Beacom, N.F. Bell, D. Hooper, J.G. Learned, S. Pakvasa
and T.J. Weiler, Phys. Rev. Lett. {\bf 92}, 011101 (2004).
\bibitem{MNS}
Z. Maki, M. Nakagawa and S. Sakata, Prog. Theor. Phys. {\bf 28},
870 (1962). 
\bibitem{gribov}
V. Gribov and B. Pontecorvo, Phys. Lett. {\bf B28}, 463 (1969).
\bibitem{FS}
K. Fujikawa and R. Shrock, Phys. Rev. Lett. {\bf 45}, 963 (1980).
\bibitem{wess}
J. Wess and J. Bagger, {\it Supersymmetry and Supergravity}
 (Princeton Univ. Press, 1992).\\
S. Weinberg, {\it The Quantum Theory of Fields}, Vol.III 
(Cambridge Univ. Press, 2000).
\bibitem{schechter}
J. Schechter and J.W.F. Valle, Phys. Rev. {\bf D22}, 2227 (1980).
\bibitem{doi}
M. Doi, T. Kotani, H. Nishiura, K. Okuda and E. Takasugi,
Phys. Lett. {\bf 102B}, 323 (1981), and references therein.
\bibitem{KM}
M. Kobayashi and T. Maskawa, Prog. Theor. Phys. {\bf 49},
652 (1973). 
\bibitem{nieves}
J.F. Nieves, Phys. Rev. {\bf D26}, 3152 (1982).
\bibitem{LM}
C.S. Lim and W.J. Marciano, Phys. Rev. {\bf D37}, 1368 (1988).
\bibitem{glashow}
J. Steinberger, as is quoted in S.L. Glashow, `` Fact and fancy 
in neutrino physics. 2'', hep-ph/0306100. 
\bibitem{aulakh}
C. Aulakh and R. Mohapatra, Phys. Lett. {\bf B119}, 136 (1983).
\bibitem{hall}
L. Hall and M. Suzuki, Nucl. Phys. {\bf B231}, 419 (1984).
\bibitem{chikashige}
Y. Chikashige, R. Mohapatra and R. Peccei, Phys. Lett. {\bf B98},
265 (1980).
\bibitem{buchmuller}
W. Buchmuller, S.T. Love, R.D. Peccei and T. Yanagida,
Phys. Lett. {\bf B115}, 233 (1982).\\
W. Buchmuller, R.D. Peccei and T. Yanagida,
Phys. Lett. {\bf B124}, 67 (1983).
\bibitem{nitta}
M. Eto, M. Nitta and N. Sakai, ``Effective theory on non-Abelian
vortices in six dimensions '', hep-th/0405161, and 
references therein.  
\end{thebibliography}
\end{document}